\documentclass[fleqn,usenatbib]{mnras}
\usepackage{newtxtext,newtxmath}
\usepackage{geometry}
\usepackage{natbib}
\usepackage{float}
\usepackage[T1]{fontenc}

\DeclareRobustCommand{\VAN}[3]{#2}
\let\VANthebibliography\thebibliography
\def\thebibliography{\DeclareRobustCommand{\VAN}[3]{##3}\VANthebibliography}

\usepackage{graphicx}	
\usepackage{amsmath}	

\title[3D 2-fluid Alfv\'en waves]{3D numerical simulations of propagating two-fluid, torsional Alfv\'en waves and heating of a partially-ionized solar chromosphere
}

\author[B.~Ku\'zma et al.]{
B.~Ku\'zma,$^{1,2}$\thanks{E-mail: blazej.kuzma@kuleuven.be}
K. Murawski,$^{2}$
and S. Poedts$^{1,2}$
\\
$^{1}$Center for Mathematical Plasma Astrophysics, Department of Mathematics, KU Leuven, Celestijnenlaan 200B, 3001 Leuven \\
$^{2}$Institute of Physics, University of Maria Curie-Sk{\l}odowska, Pl. M. Curie-Sk{\l}odowskiej 5, 20-031 Lublin, Poland
}

\date{Accepted XXX. Received YYY; in original form ZZZ}

\pubyear{2020}

\begin{document}
\label{firstpage}
\pagerange{\pageref{firstpage}--\pageref{lastpage}}
\maketitle

\begin{abstract}
We present a new insight into the propagation, attenuation and dissipation of two-fluid, torsional Alfv\'en waves in the 
context of heating of the lower solar atmosphere. 
By means of numerical simulations of the partially-ionized plasma, we solve the set of two-fluid equations for ion plus electron and neutral fluids in three-dimensional (3D) Cartesian geometry. We implement initially a current-free magnetic field configuration, corresponding to a magnetic flux-tube that is rooted in the solar photosphere and expands into the chromosphere and corona. We put the lower boundary of our simulation region in the low chromosphere, where ions and neutrals begin to decouple, and implement there 
a monochromatic driver that directly generates Alfv\'en waves with a wave period of 30$\;$s. 
As the ion-neutral drift increases with height, the two-fluid effects become more significant and the energy carried by both Alfv\'en and magneto-acoustic waves can be thermalized in the process of ion-neutral collisions there. In fact, we observe a significant increase in plasma temperature along the magnetic flux-tube. In conclusion, the two-fluid torsional Alfv\'en waves 
can potentially play a role in the heating of
the solar chromosphere. 
\end{abstract}


%
\section{Introduction}
To understand the multitude of physical processes occurring on the Sun, one has to know how 
the solar plasma dynamics is determined by the solar magnetic field. One of the omnipresent magnetic structures in the solar photosphere are magnetic flux-tubes. Observed usually in the photosphere and low  chromosphere \citep[e.g.,][]{Hansteen2006,DePontieu2007,Voort2009,Kuridze2015,Srivastava2017} and rooted in between granules, they are prone to  highly turbulent motions associated with the solar granulation. The latter is an efficient source of waves that are propagating into the solar atmosphere, and some of which were interpreted as magnetohydrodynamic-gravity waves \citep[e.g.,][]{Musielak2001}. 
Recent high-resolution observations revealed that the solar atmosphere is indeed permeated by diversity of waves, which in some cases carries energy-flux sufficient to heat the atmospheric plasma 
\citep[e.g.,][]{Jess2009,DePontieu2014,Morton2015,Jafarzadeh2017,Srivastava2017}. 
In the light of these discoveries, wave heating is widely discussed as a potential solution of the coronal heating problem. The energy transport from the photosphere, through the chromosphere and into the corona 
strongly depends on the efficiency of wave reflection from the inhomogeneous regions \citep{Hollweg1982, Cranmer2005}. In particular, the mentioned omnipresent solar magnetic flux-tubes may act as guidelines for Alfv\'en waves, which were theoretically predicted by \cite{Alfven1942}, and have been potentially observed in the solar atmosphere \citep[e.g.,][]{Tomczyk2007,Srivastava2017}. The linear Alfv\'en waves propagation in the isothermal solar atmosphere was investigated numerically and analytically and cut-off frequencies were revealed by \cite{Perera2015}. It was also shown that a finely-structured flux-tube can be a guide not only for Alfv\'en waves, but also for fast  magneto-acoustic kink waves \citep{Kuzma2018} and localized Alfv\'en pulses may play potential role in vertical mass transport and formation of lower solar atmospheric jets \citep{Scalisi2021}. However, the actual mechanisms of energy dissipation, and thus its efficiency and  ability to compensate the radiative losses, is still a matter of ongoing discussion. 

The many different types of waves have complex interactions: mode-coupling can transfer energy from one mode into another, the waves refract while propagating through the inhomogeneous atmosphere, also their polarization changes as they propagate, and real physical absorption and conversion of the wave power into heat can take place on top of all of those other processes \citep{Khomenko2018}. 
Neutrals can possibly play an important role in photosphere and chromosphere, as plasma is only partially ionized there 
\citep[e.g.,][]{Zaqarashvili2011}. The collisions between ions and neutrals can lead, among other things, to wave damping and energy dissipation \citep[e.g.,][]{Piddington1956,Watanabe1961,Kulsrud1969,Haerendel1992,DePontieu1998,James2002,Erdelyi2004,Khodachenko2004,Forteza2007}. It was demonstrated that also the properties of Alfv\'en waves are prone to partial ionization \citep{Soler2013}. The above mentioned collisional energy dissipation and as a result collisional heating should be taken into consideration in the context of the ongoing discussion on heating of the chromosphere. Thus, the two-fluid processes are essential for a correct description of the wave-processes occurring in the lower layers of the solar atmosphere.

The chromospheric heating by Alfv\'en waves due to two-fluid effects, mainly ion-neutral collisions, was first discussed in the pioneering paper of \cite{Piddington1956} who considered waves propagating in a partially-ionised gas with ionized plasma and neutral gas treated as separated, independently moving fluids. 
He estimated the frictional heating in the chromosphere as $10^5\;$erg cm$^{-2}$ s$^{-1}$. 
Later, \cite{Osterbrock1961} argued this value to be overestimated, pointing out that the effective cross-section for ion-neutral collisions is larger and the real frequency spectrum is considerably lower and the frictional damping-heating rate is in fact very small. \cite{Haerendel1992} discussed weakly damped Alfv\'en waves, due to ion-neutral collisions, as drivers of solar chromospheric spicules. \cite{Goodman2011} 
investigated the effect of Pedersen current dissipation of driven in the photosphere Alfv\'en oscillations on the chromospheric heating rate.
He showed that the latter is comparable with the observationally estimated radiative losses of $10^7\;$erg cm$^{-2}$ s$^{-1}$ in the chromosphere.  \cite{Zaqarashvili2013}, using a single-fluid magnetohydrodynamic (MHD) model supplemented by a Cowling diffusion coefficient in the presence of neutral helium, showed strong damping of short-period torsional Alfv\'en waves by ion-neutral collisions. 
The propagation of Alfv\'en waves, and their collisional dissipation and heating in a 1.5D partially ionised solar chromosphere, 
was studied by \cite{DePontieu2001}, \cite{Leake2005}, and \cite{Tu2013}. 
Additionally, \cite{Arber2016} using a 1.5D non-ideal MHD modelling, demonstrated that it is possible to heat the chromospheric plasma by direct resistive dissipation of high-frequency Alfv\'en waves through Pedersen resistivity. \cite{Shelyag2016} modelled non-linear wave propagation in a magnetic flux tube
with the effect of partial ionization taken into account by the ambipolar term from the generalized Ohm’s law in a single-fluid quasi-MHD regime. They showed, that ambipolar diffusion can lead to effective dissipation of MHD waves perturbation and, as result, to temperature increase in chromospheric magnetic structures. 
\cite{Soler2017}, and \cite{Soler2019} investigated energy transport and heating by torsional Alfv\'en waves propagating along magnetic flux-tubes expanding from the photosphere to the low corona, with the assumption of stationary state propagation. They found that waves with wave periods shorter than $20\;$s are dissipated in the chromosphere, and thereby the energy sufficient to balance the radiative losses is thermalized there. They also demonstrated that in the photosphere heating through Ohmic dissipation is dominating over heating through ion-neutral collisions. 

For a long time it was a subject of intensive discussions whether two-fluid effects play any role in the low frequency regime \citep[e.g.,][among others]{Soler2013,Ballai2019}. Recently published numerical results strongly suggest that ion-neutral collisions can affect the transport of energy and its dissipation in a partially-ionized chromosphere. For instance, \cite{Soler2019}, using a photospheric broadband driver that contains a spectrum of frequencies corresponding to wave periods ranging from $3.33\;$s up to $2.78\;$h, investigated plasma heating and energy transfer by two-fluid Alfv\'en waves. \cite{Popescu2019} showed  collisional damping and energy dissipation for monochromatic waves with wave periods starting from $1\;$s up to $20\;$s. Moreover, in the recent works on linear and nonlinear two-fluid Alfv\'en waves \cite{MartnezGmez2018} and \cite{Kuzma2020} revealed a rise of the plasma temperature in the solar atmosphere resulting from the collisional heating. 
The two-fluid effects also play an important role when non-linear effects become important, particularly in the context of wave damping and plasma heating \citep[e.g.,][]{Kuzma2019,Wojcik2020,Fan2020}. On top of that, \cite{Murawski2020} demonstrated that ion-neutral collisions may solely solve the heating problem of a quiet region of the chromosphere.
 
The main goal of the present paper is to investigate the impact of two-fluid effects on (a) propagation, attenuation and dissipation of torsional Alfv\'en waves, (b) chromospheric heating, and (c) energy transfer to the upper layers of the solar corona. 
We aim to follow and complement both the previous studies of \cite{Popescu2019} and \cite{Kuzma2019} by taking into account a 3D model of the solar atmosphere, and the study of \cite{Soler2019} by taking into account a non-static, time-varying model of propagating Alfv\'en waves. As we want to  focus solely on two-fluid effects on wave propagation and plasma heating, we limited our model by not taking into account non-ideal terms in the induction equation, nor ionization and recombination \citep{Popescu2019} and resistivity \citep{Soler2019}. 
Ultimately, we want to determine and stress the importance of two-fluid effects for processes with time-scales longer than the characteristic time-scales of ion-neutral collisions. 

This paper is organized as follows. In Section~2, we present the two-fluid equations for ions plus electrons and neutrals. In Section~3, we describe the 3D magneto-hydrostatic model of a partially-ionized solar atmosphere and present results of numerical simulations. Finally, we conclude and summarize our work in Section~4. 

%
\section{Two-fluid plasma model}
To model the lower atmospheric layers of the Sun, we consider a gravitationally stratified and magnetically confined plasma that consists of two components: 
electric charges (ions plus electrons) and neutral fluid (neutrals) that interact via ion-neutral collisions. 
The non-ideal terms in the induction equation, ionization and recombination \citep{Popescu2019} as well as resistivity   \citep{Soler2019} 
are not considered in the present model. This model is governed by the equations describing mass conservation:
\begin{align}
\frac{\partial \varrho_{\rm n}}{\partial t}+\nabla\cdot(\varrho_{\rm n} \mathbf{V_{\rm n}}) &= 0 \, ,\\
\frac{\partial \varrho{\rm {\rm _i}}}{\partial t}+\nabla\cdot(\varrho{\rm {\rm _i}} \mathbf{V_{\rm i}}) &= 0 \, ; 
\end{align}
the momentum equations: 
\begin{equation}
\begin{split}
& \varrho_{\rm n}\frac{\partial  \mathbf{V_{\rm n}}}{\partial t}+\varrho_{\rm n} (\mathbf{V_{\rm n}}\cdot \nabla)\mathbf{V_{\rm n}} = 
-\nabla p_{\rm n} + \varrho_{\rm n} \mathbf{g} - \alpha_{\rm c} (\mathbf{V_{\rm n}}-\mathbf{V_{\rm i}} ) \, , 
\end{split}
\end{equation}
\begin{equation}
\begin{split}
& \varrho{\rm {\rm _i}}\frac{\partial  \mathbf{V_{\rm i}}}{\partial t}+\varrho{\rm {\rm _i}} (\mathbf{V_{\rm i}}\cdot \nabla)\mathbf{V_{\rm i}} = \\ & -\nabla p{\rm {\rm _{i}}} + \frac{1}{\mu} (\nabla \times \mathbf{B})\times\mathbf{B} +\varrho{\rm {\rm _i}} \mathbf{g} + \alpha_{\rm c} (\mathbf{V_{\rm n}}-\mathbf{V_{\rm i}}) \, ; 
 \end{split}
 \end{equation}
and the energy equations:
\begin{equation}
\begin{split}
 & \frac{\partial E_{\rm n}}{\partial t}+\nabla\cdot((E_{\rm n}+p_{\rm n})\mathbf{V_{\rm n}}) =  Q_{\rm n} +(\varrho_{\rm n} \mathbf{g} - \alpha_{\rm c} (\mathbf{V_{\rm n}}-\mathbf{V_{\rm i}} )) \cdot \mathbf{V_{\rm n}} \, , \\
 \end{split}
\end{equation}
\begin{equation}
\begin{split}
& \frac{\partial E{\rm {\rm _i}}}{\partial t}+\nabla\cdot \left[\left(E{\rm {\rm _i}}+p{\rm {\rm _{i}}}+\frac{{\mathbf{B}}^2}{2\mu}\right)\mathbf{V_{\rm i}}-\frac{\mathbf{B}}{\mu}(\mathbf{V_{\rm i}} \cdot {\mathbf{B}})\right]  = \\
 &  Q{\rm {\rm _i}} +(\varrho{\rm {\rm _i}} \mathbf{g} + \alpha_{\rm c} (\mathbf{V_{\rm n}}-\mathbf{V_{\rm i}})) \cdot \mathbf{V_{\rm i}} \, ; 
\end{split}
\end{equation}
which are supplemented by induction equation and the solenoidal condition:
\begin{align}
 \frac{\partial \mathbf{B}}{\partial t} = \nabla \times (\mathbf{V_{\rm i}} \times \mathbf{B})\,,\hspace{3mm}  \nabla \cdot \mathbf{B} &= 0 \, .
\end{align}
Here, 
%
%
\begin{align}
E_{\rm n} = \frac{p_{\rm n}}{\gamma-1} + \frac{1}{2}\varrho_n |\mathbf{V_{\rm n}}|^2\, ,
\end{align}
and
\begin{align}
E_{\rm i} = \frac{p_{\rm i}}{\gamma-1} + \frac{1}{2}\varrho{\rm {\rm _i}} |\mathbf{V{\rm _i}}|^2 + \frac{ |\mathbf{B}|^2}{2\mu},
\end{align}
where subscripts ${\rm _n}$ and ${\rm _i}$ denotes ion and neutral components of energy densities, $E_{\rm n}$ ($E_{\rm i}$), mass densities, $\varrho_{\rm n}$ ($\varrho{\rm _i}$), and temperatures, $T_{\rm n}$ ($T_{\rm i}$). The gas pressure is indicated for neutrals by $p_{\rm n}$, and for ions plus electrons by $p_{\rm i}$. The constant $\mu$ is the magnetic permeability of the medium, $\mathbf{B}$ denotes the magnetic field, and $\gamma=5/3$ the adiabatic index. The gravitational acceleration vector is directed towards negative $y$-axis in the Cartesian coordinate system we consider and its magnitude is $g = 274.78 \rm ~m~s^{-2}$. 

Note, that in our model we consider hydrogen as the main plasma component. The influence from heavier elements is taken from the OPAL solar abundances repository (e.g., \citealp{Vogler2005}). 
The mean masses of ions plus electrons and neutrals are $\mu_{\rm i}=0.58$ and $\mu_{\rm n}=1.21$, respectively. The above-described fluids interact via ion-neutral collisions. The effect of the interaction between these fluids depends on the ion-neutral friction coefficient. 
Following \cite{Zaqarashvili2011}, we assume that 
$\alpha_{\rm in}=\alpha_{\rm ni}=\alpha_{\rm c}$, 
and with use of formula derived by \cite{Braginskii1965} we find: 
%
\begin{equation}
\alpha_{\rm c} = \frac{4}{3} \frac{\sigma_{\rm in}\,\varrho_{\rm i}\,\varrho_{\rm n}}{m_H(\mu_{\rm i}+\mu_{\rm n})} \sqrt{\frac{8k_B}{\pi m_H} \left(\frac{T{\rm _i}}{\mu_{\rm i}}+\frac{T_{\rm n}}{\mu_{\rm n}}\right)} \, .
\end{equation}
Here, $m_{\rm H}$ corresponds to the hydrogen mass, $k_B$ is Boltzmann's constant, and $\sigma_{\rm in}$ is the ion-neutral collision cross-section taken as a quantum value (\citealp{Vranjes2013}). 
Note, that the frictional interaction between these fluids results in additional heat production and exchange terms proportional, respectively, to the square of the velocity difference and the temperature difference \citep[e.g.,][and references cited therein]{Ballester2018}: 
\begin{align}
Q_{\rm n} = \alpha_{c} \left[ \frac{1}{2} \vert\mathbf{V{\rm _i}}-\mathbf{V_{\rm n}}\vert ^2 - \frac{3}{2} \frac{k_B}{m_{\rm H}(\mu_{\rm i}+\mu_{\rm n})} (T_{\rm n}-T{\rm _i}) \right]\,, \\
Q{\rm _i} = \alpha_{c} \left[ \frac{1}{2} \vert\mathbf{V{\rm _i}}-\mathbf{V_{\rm n}}\vert ^2 - \frac{3}{2} \frac{k_B}{m_{\rm H}(\mu_{\rm i}+\mu_{\rm n})} (T{\rm _i}-T_{\rm n}) \right]\,.
\end{align}
This frictional heating is a two-fluid effect, however it can be studied in single-fluid MHD models that include the ambipolar term in the induction equation and the corresponding ambipolar heating term in the energy equation when the two fluids are strongly coupled. In this case, the frictional heating term consistently reverts to the ambipolar heating term in the single-fluid approximation \citep{Khomenko2014,Ballester2018}. We study its impact on partially-ionized plasma behaviour and contribution to overall heating within a two-fluid model. From Eqs.~(11)-(12) we infer that collisional heating arises when the mass densities of both fluids are similar and both fluids are still coupled (thus non-zero $\alpha_{\rm c}$), but this coupling is weak enough to allow ion-neutral drift, $\mathbf{V{\rm _i}}-\mathbf{V_{\rm n}}$, attaining non-zero values. We infer that these conditions are met in the chromosphere. %

\begin{figure}
	\begin{center}
		\mbox{
		 \includegraphics[scale=0.47]{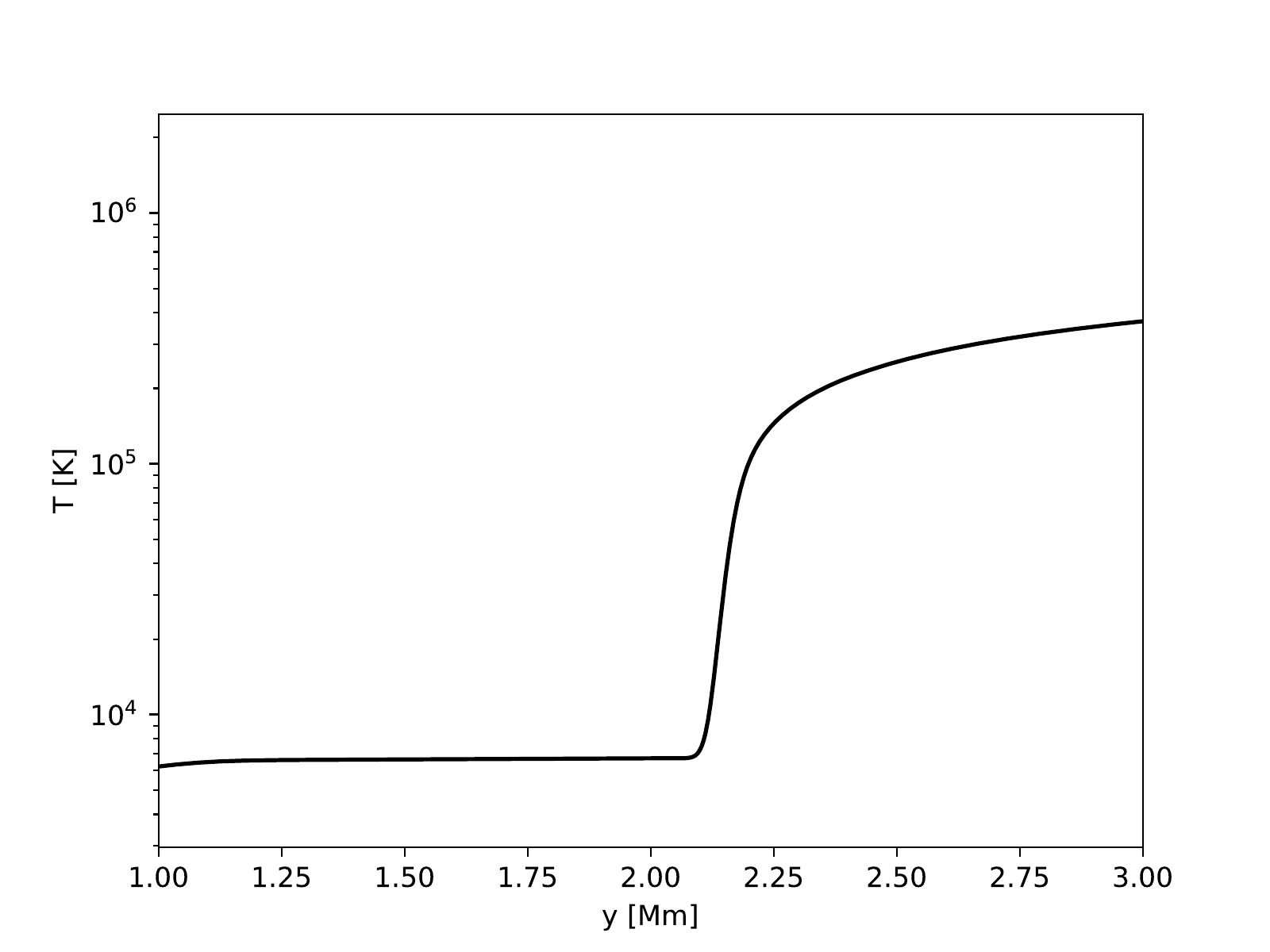}} \\
		 			\hspace{-0.0cm}
		 \mbox{			
		 \includegraphics[scale=0.47]{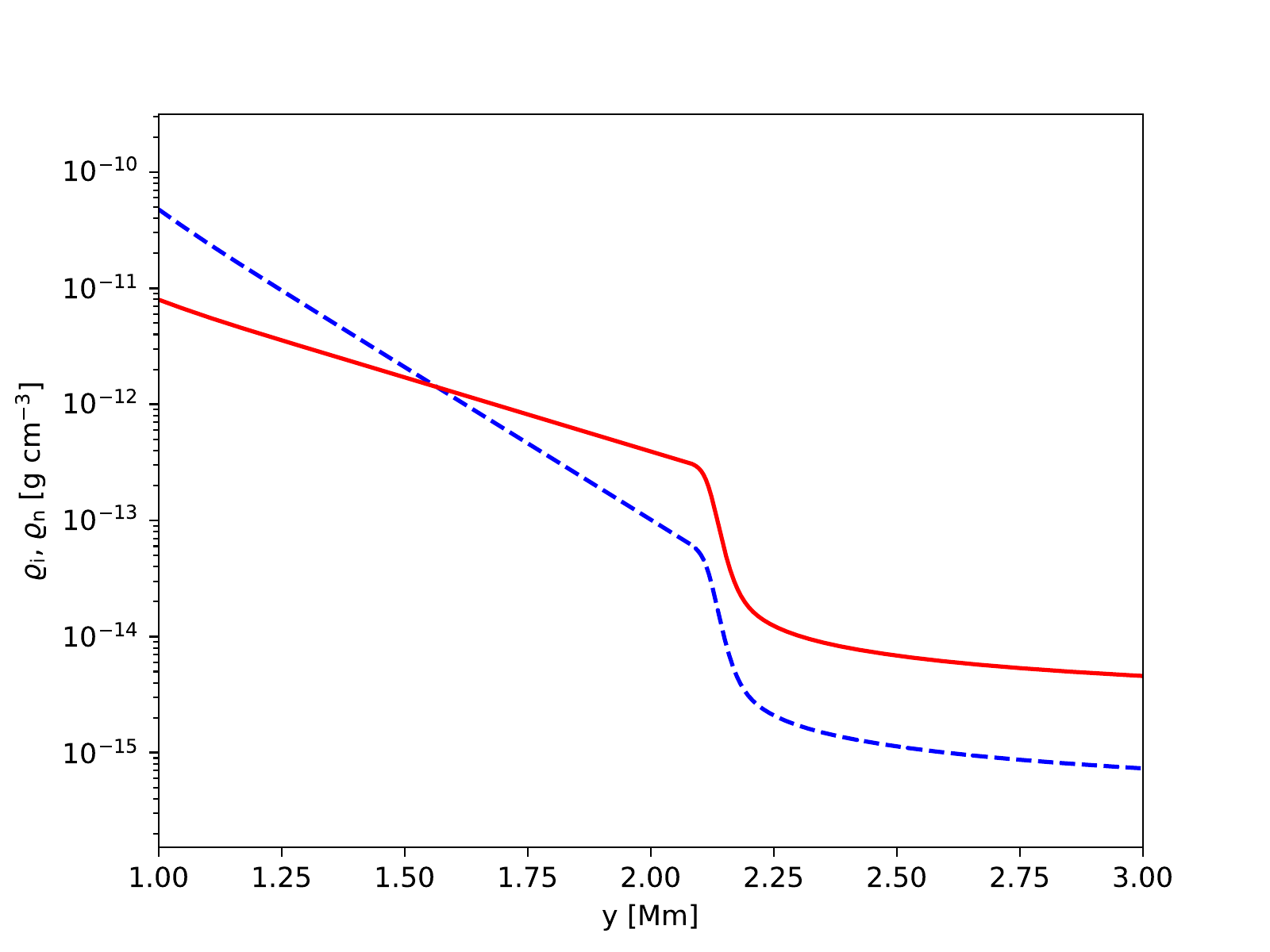}} \\
		 			\hspace{-0.0cm}
		 \mbox{			
		 \includegraphics[scale=0.47]{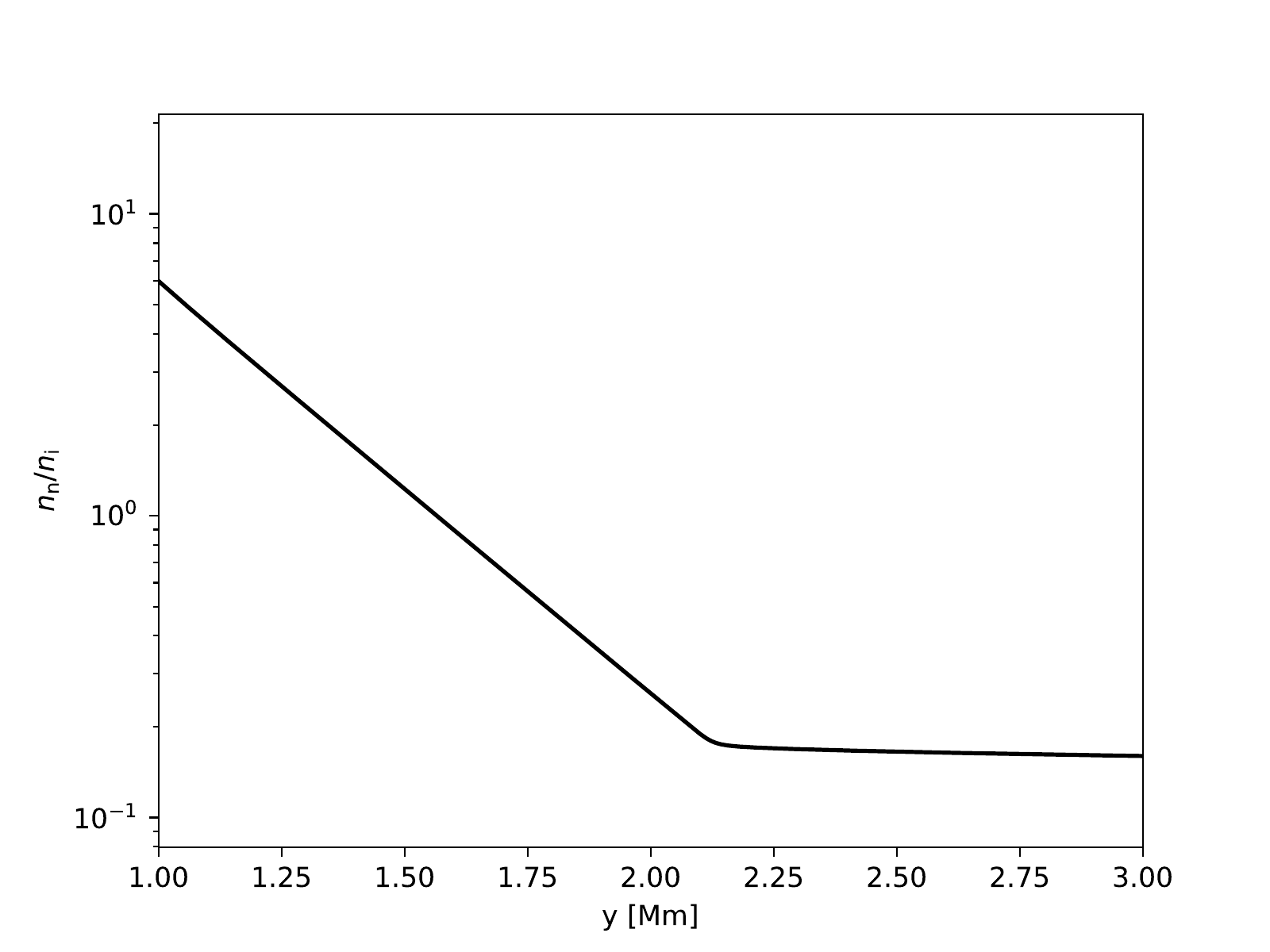}
	     }
		\caption{Vertical profiles of temperature $T$ (top), corresponding ion $\varrho_{\rm i}$ (solid line, middle) and neutral $\varrho_{\rm  n}$ (dashed line, middle) mass densities, and the ionization ratio $n_{\rm n}/n_{\rm i}$ (bottom).}
		\label{fig:solar_profiles2}
	\end{center}
\end{figure}
\section{Numerical simulations of 3D Alfv\'en waves}
\begin{figure}
	\begin{center}
	  			\vspace{-1.5cm}
		\mbox{
					\hspace{-0.5cm}
		 \includegraphics[scale=0.45]{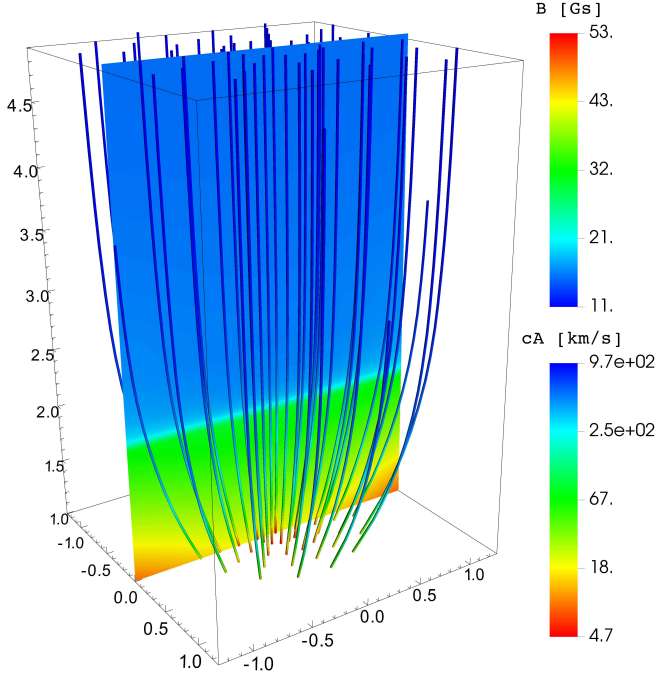}
	   }
	  			\vspace{-1.5cm}
		\caption{The equilibrium magnetic field lines with 
		magnetic field strength, $B$,  
		expressed in Gauss and 
		the equilibrium 
		Alfv\'en speed profile, $c_A$, expressed in km s$^{-1}$. 
		}
		\label{fig:solar_profiles}
	\end{center}
\end{figure}
\subsection{Hydrostatic equilibrium model of the solar atmosphere}
In our model, the solar atmosphere is assumed to be initially 
(at $t=0\;$s) in a
hydrostatic (${\bf V_{\rm i,n}}={\bf 0}$) equilibrium. 
%
Taking the ideal gas law for ion and neutral fluids into account: 
\begin{equation}
p_{\rm n}=\frac{k_{\rm B}}{m_{\rm H}\mu_{\rm n}}\varrho_{\rm n} T_{\rm n}\,,\qquad\hbox{and}\qquad p{\rm {\rm _{i}}} = \frac{k_{\rm B}}{m_{\rm H}\mu_{\rm i}}\varrho{\rm {\rm _i}} T{\rm {\rm _i}}\, ,
\end{equation}
and from the $y-$components of hydrostatic ($-\nabla p_{\rm i, \, n}+\varrho_{\rm i, \, n}\,{\mathbf{g}}=\mathbf{0}$) equations, 
we derive the equilibrium gas pressures: 
\begin{align}
p_{\rm n}(y)&=p_{0\, {\rm n}} \, {\rm exp} \left[-\int^{y}_{y_{\rm ref}}\frac{dy}{\Lambda_{\rm n}(y)} \right] \, , \hspace{3mm} \\
p_{\rm i}(y)&=p_{0\, {\rm i}} \, {\rm exp} \left[ - \int^{y}_{y_{\rm ref}}\frac{dy}{\Lambda_{\rm i}(y)} \right] \, .
\end{align}
Here, $p_{0\, {\rm n}}=3\cdot 10^{-3}\;$dyn cm$^{-2}$ ($p_{0\, {\rm i}}=10^{-1}\;$dyn cm$^{-2}$) corresponds to the pressure of neutrals (ions) at the reference level, $y_{\rm ref}=50\;$Mm, and  
\begin{equation}
\Lambda_{\rm n}=\frac{k_{\rm B} \, T(y)}{m_{\rm H}\,\mu_{\rm n} \, g}\,,\qquad\hbox{and}\qquad \Lambda_{\rm i}=\frac{k_{\rm B} \, T(y)}{m_{\rm H}\,\mu_{\rm i} \, g}\, ,
\end{equation}
are, the neutral (ion) pressure scale-heights, respectively.
Note that the ion and neutral mass density profiles depend solely on the temperature (Fig.~1, top), which is taken from the semi-empirical model of \cite{Avrett2008}. As a result of differences between these profiles (Fig.~1, middle), the ionization degree depends on height above the photosphere. Starting from the weakly ionized bottom of the chromosphere, where the neutral mass density exceeds the ion mass density by about an order of magnitude, through ion-neutral equilibrium at about $900\;$km below the located at $y\approx2.1\;$Mm transition region, to the one million hot solar corona which, as the mass density of neutrals experiences a sudden fall-off with height, is essentially fully ionized (Fig.~1, bottom). 

%
\begin{figure*}
	\begin{center}
		\mbox{
			\hspace{-0.5cm}
		 \includegraphics[scale=0.5]{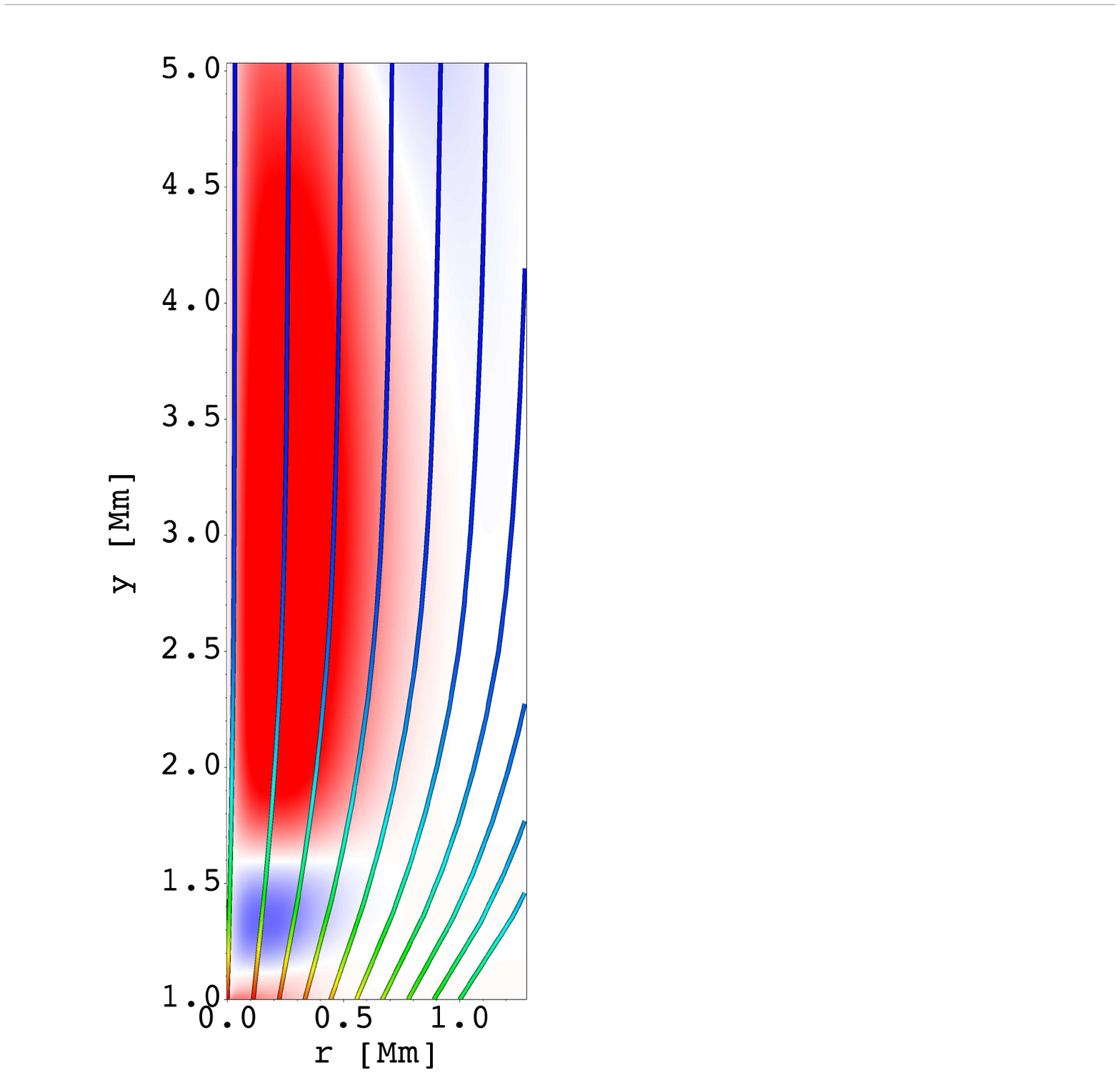} 
		 			\hspace{-6.0cm}
		 \includegraphics[scale=0.5]{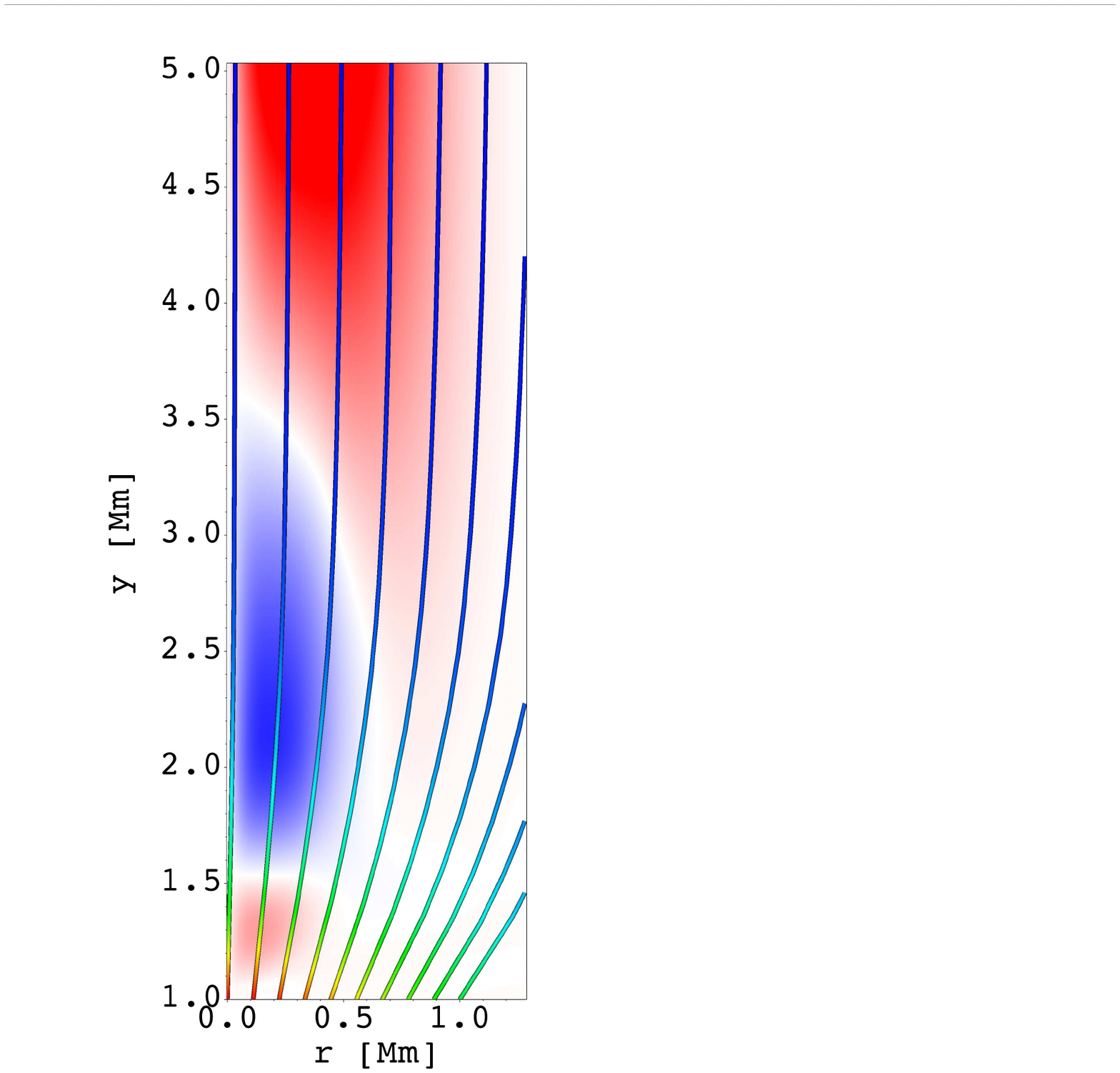} 
		 			\hspace{-6.0cm}
		 \includegraphics[scale=0.5]{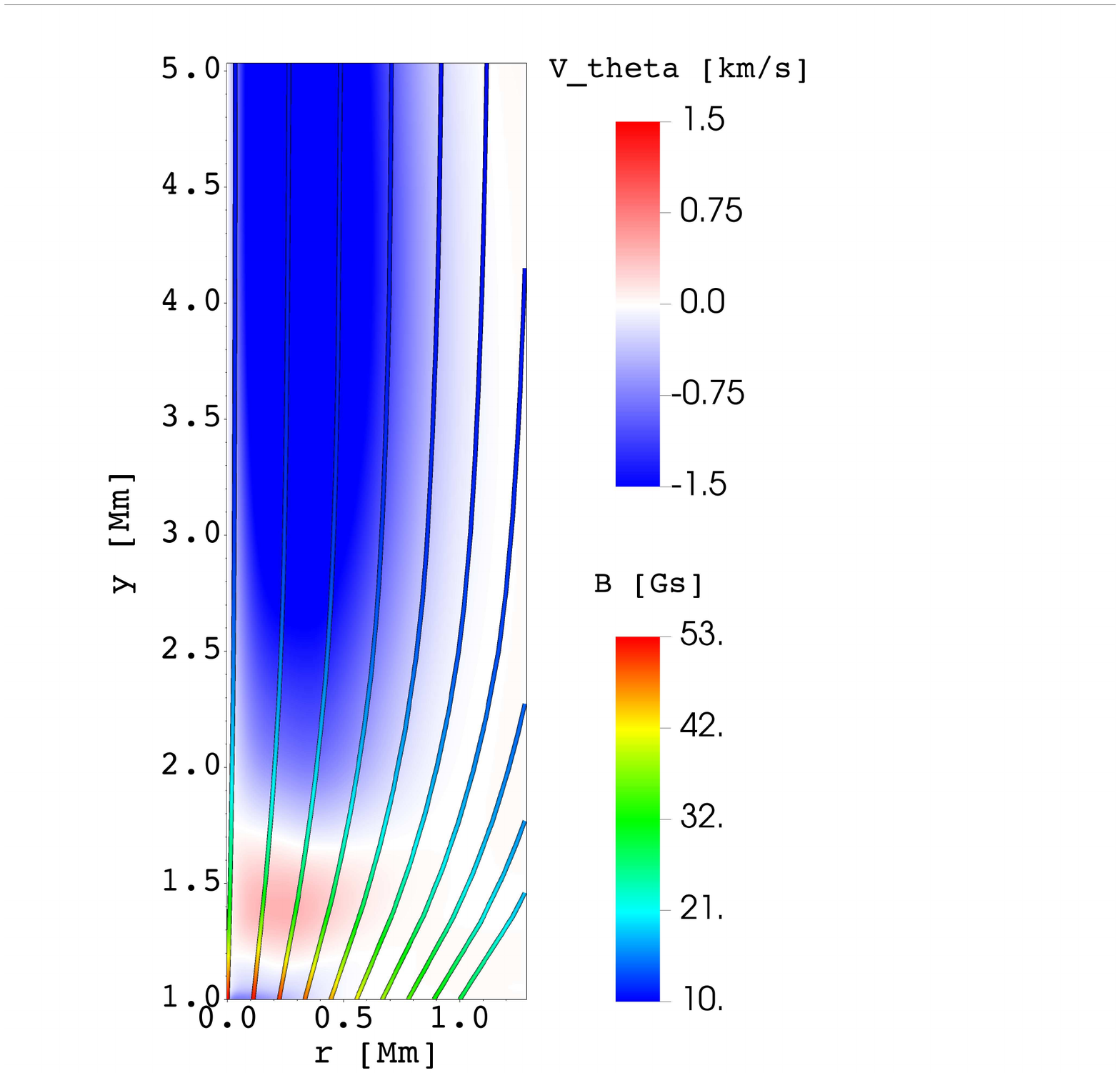}
	    }
	    	  			\vspace{-2.2cm}
		\caption{Three spatial profiles of the azimuthal 
		ion velocity, $V_{{\rm i}\, \theta}(r,y)$, with $r=\sqrt{x^2+z^2}$, at $t = 980\;$s, $t = 990\;$s, and $t = 1000\;$s (from left to right) overlayed by magnetic field lines, 
		the case of 
		$P_{\rm d}=30\;$s and 
		$A_{V}=1\;$km s$^{-1}$.}
		\label{fig:solar_profiles2}
	\end{center}
\end{figure*}

In the above described equilibrium model, we implement a current-free, and thus force-free, magnetic field with the configuration of a magnetic flux-tube. In 3D Cartesian coordinates this structure is described as (\citealp{Low1985}) 
\begin{equation}
\begin{split}
& [B_x,B_y,B_z] = \\
& \frac{S[-3x(y-a),x^2-2(y-a)^2+z^2,-3(y-a)z]}{(x^2+(y-a)^2+z^2)^{5/2}} \\
&+[0,B_{\rm V},0]\, ,
\end{split}
\end{equation}
where the magnetic field singularity is located at $y=a$, the free parameter $S$ determines its strength 
and $B_{\rm v}=10\;$G is additional vertical component of magnetic field. As a result, the described magnetic structure mimics a 
chromospheric magnetic flux-tube with strength of its magnetic field equals to $B=53\;$G at $y=1$ Mm above the photosphere, 
extending up into the corona, and attaining a magnetic field strength of $B=10\;$G at height of $y=5\;$Mm, with magnetic lines becoming 
essentially vertical at this level (Fig.~2, solid lines). When extrapolated to the photosphere, the investigated flux-tube will be rooted in-between the granules with a magnetic field strength of $B=850\;$G at its foot-point.
The spatial profile of the corresponding Alfv\'en speed, 
$c_{\rm A}(x,y)=B/\sqrt{\mu(\varrho_{\rm i}+\varrho_{\rm n})}$, 
is illustrated in Fig.~2 (colormap). In the lower chromosphere $c_{\rm A}$ attains values lower than $10\;$km s$^{-1}$ outside the flux-tube, while in its center, given by $x=z=0$ Mm it attains value of $20\;$km s$^{-1}$. 
As a result of decreasing mass density it rises higher up exceeding value of $50\;$km s$^{-1}$ and experiencing sudden jump to $900\;$km s$^{-1}$ at the transition region. 
%

%
\begin{figure} 	    	  	
	\begin{center} 	    	  			
		\mbox{	 			 		
		\vspace{-1.5cm}
	 \includegraphics[scale=0.4]{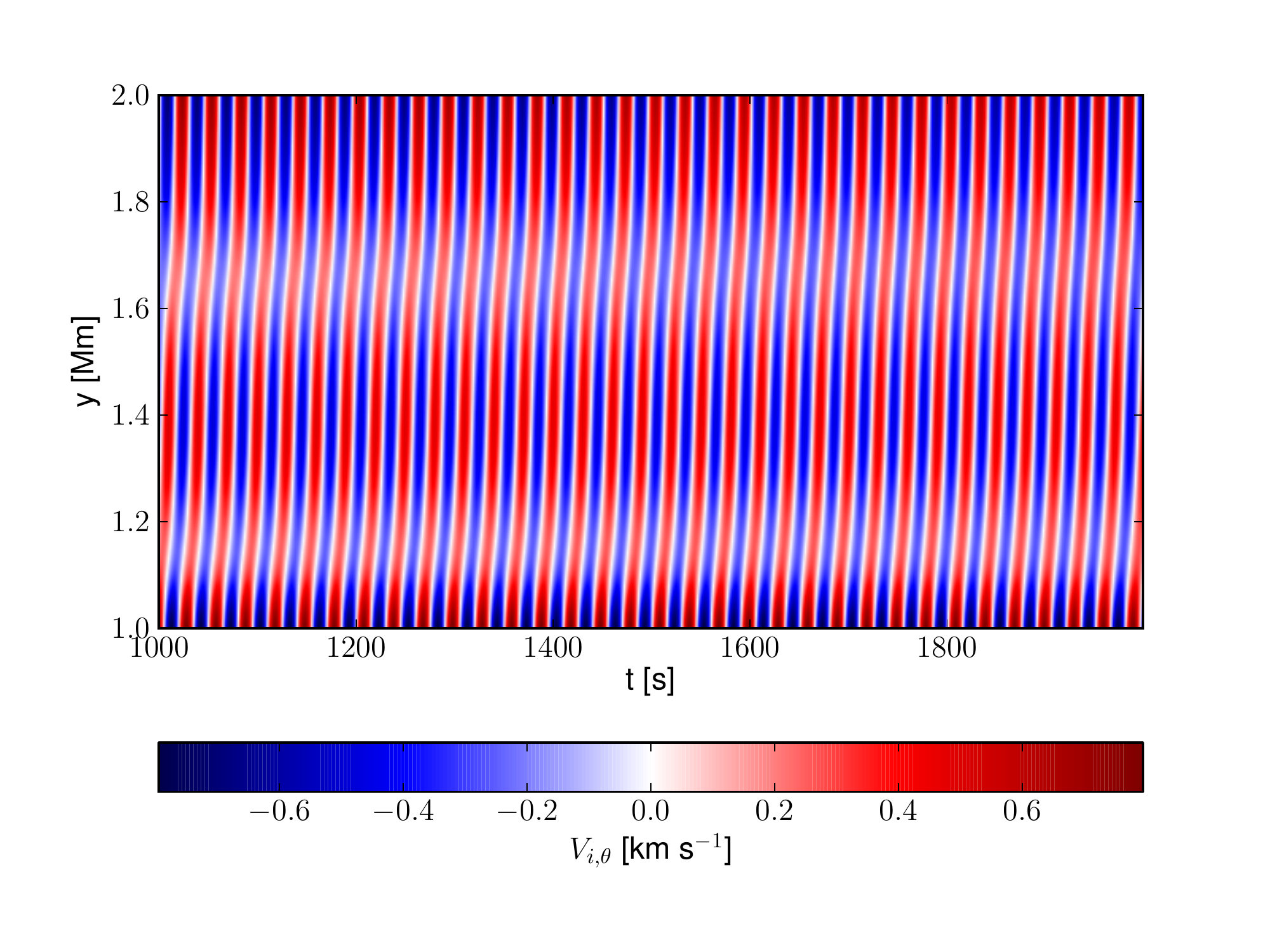} 
}	   	    	 \\
	 		\mbox{	 
	 \includegraphics[scale=0.4]{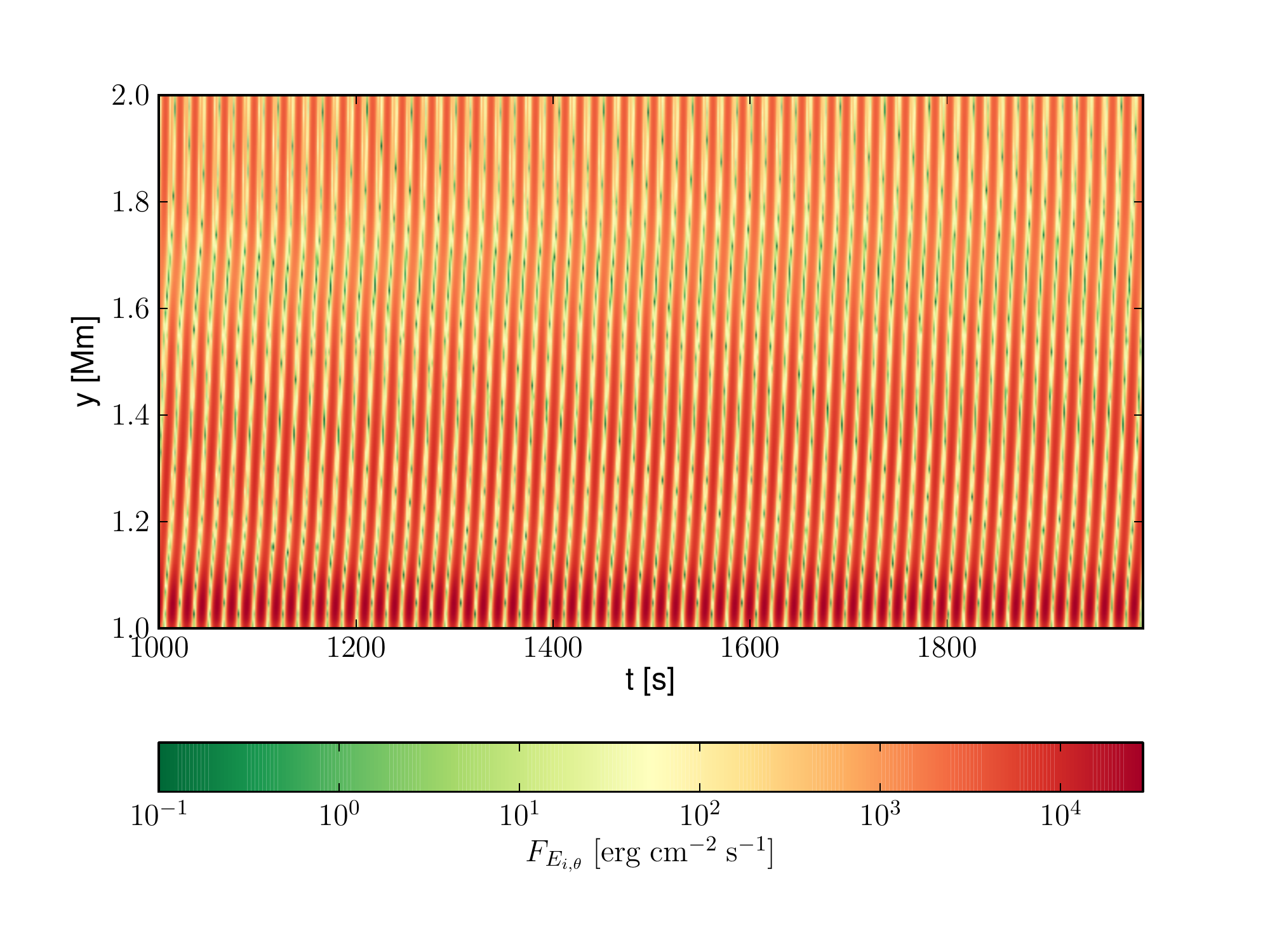}
} \\
	 		\mbox{
	 \includegraphics[scale=0.4]{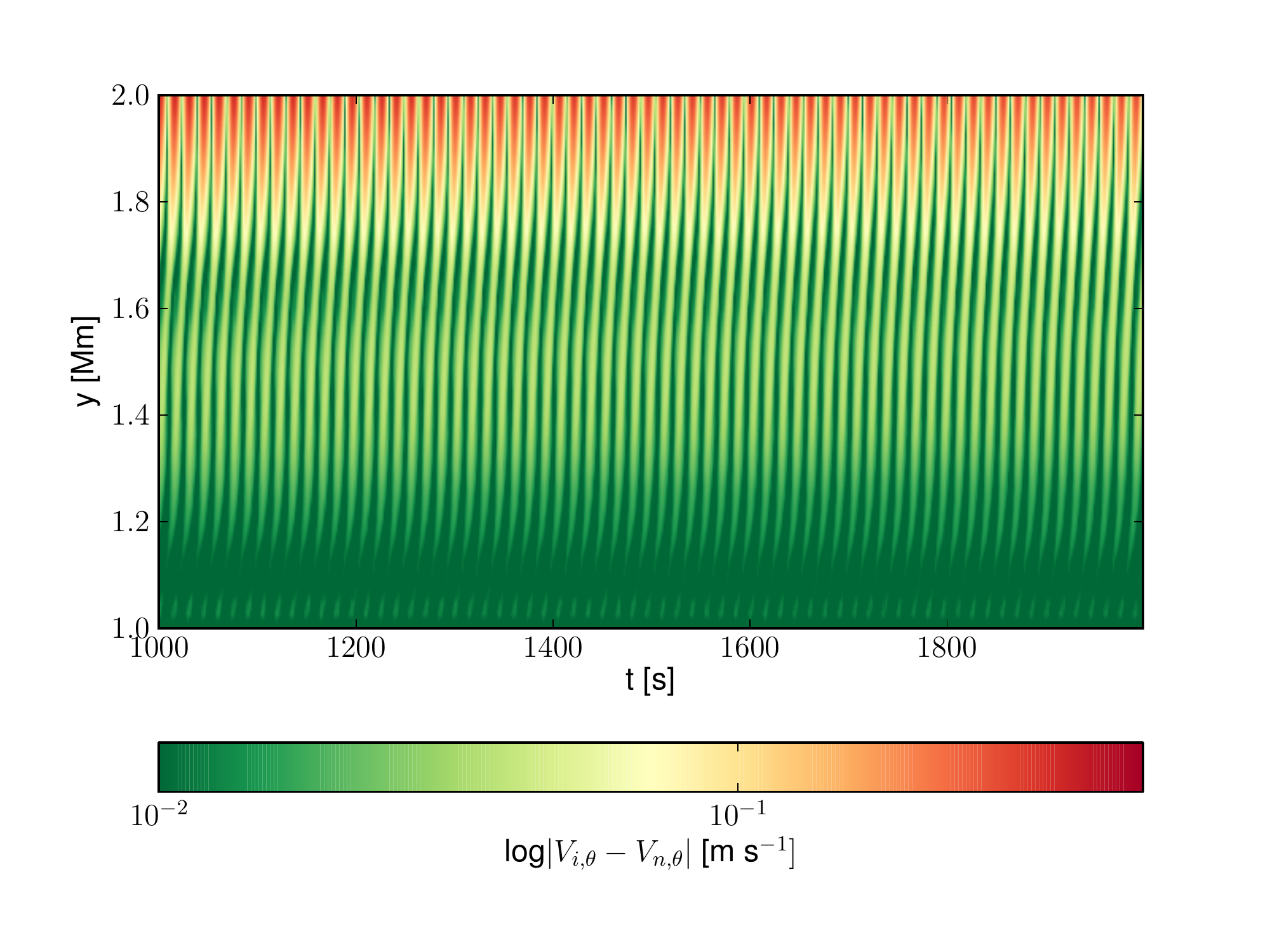}
}

		\caption{Time-distance plots for azimuthal components of ion velocity, $V_{{\rm i}\, \theta}(r=0.1\;$Mm), ion energy flux, $F_{E_{\rm i}\, \theta}(r=0.1\;$Mm), and ion-neutral drift, $\log|V_{{\rm i}\, \theta}-V_{{\rm n}\, \theta}|(r=0.1\;$Mm), from top to bottom.  
		}
	\end{center}
\end{figure}
\begin{figure}
	\begin{center}
		\mbox{
		 \includegraphics[scale=0.43]{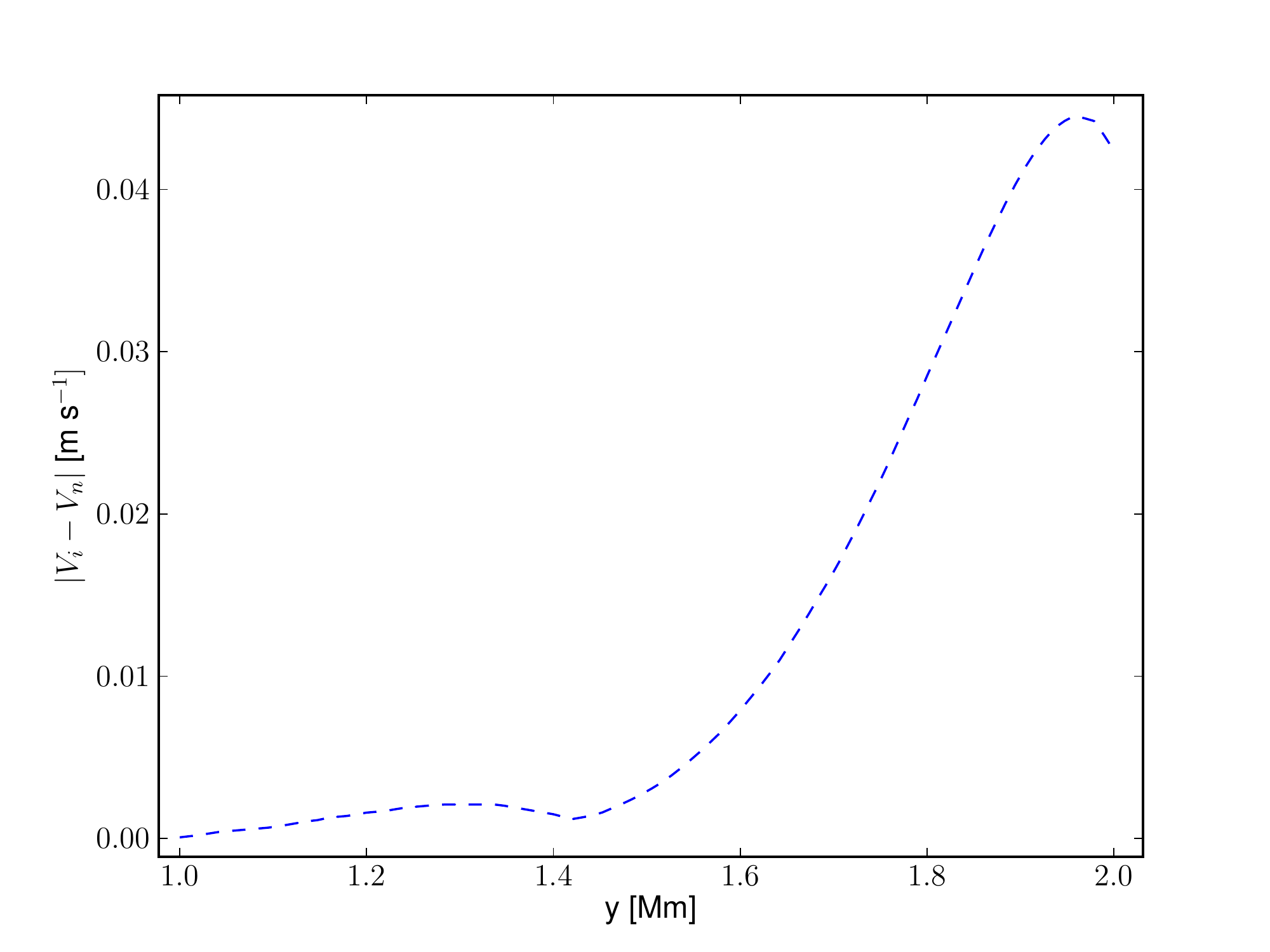}}\\
		\mbox{		 
		 \includegraphics[scale=0.43]{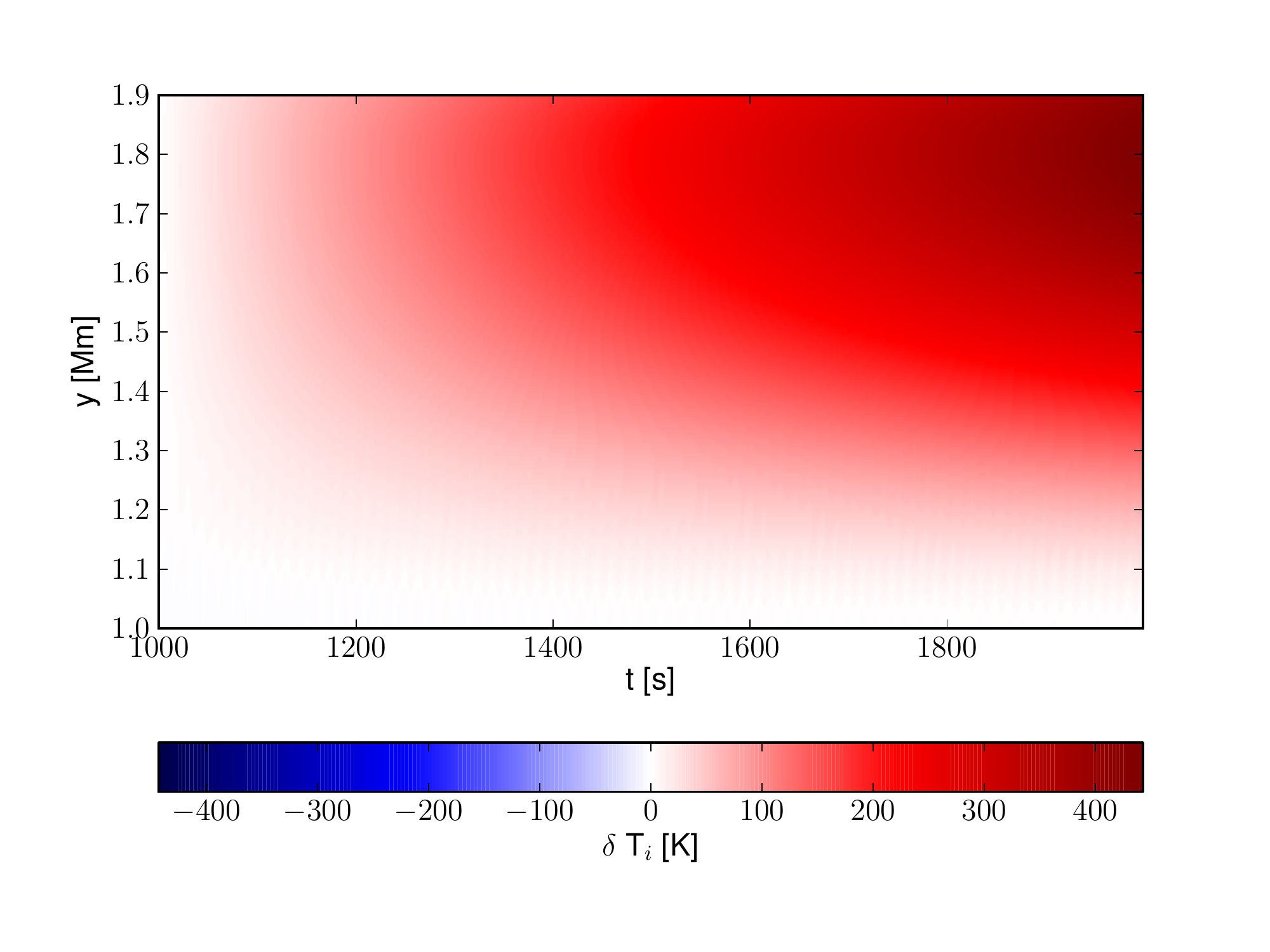}}
		\caption{Top: The maximum value of ion-neutral drift, ${V_{\rm i}-V_{\rm n}}$ carried by Alfv\'en waves. 
		Bottom: time-distance plot for relative ion temperature, $\delta T_{\rm i}$.} 
		\label{fig:solar_profiles}
	\end{center}
\end{figure}

\subsection{Numerical box and perturbations}

To solve the two-fluid equations numerically, we use the JOANNA code \citep{Wojcik2018a,Wojcik2019}. For the considered problem, we set the Courant–Friedrichs–Lewy number equal to $0.3$ and choose the TVDLF approximate Riemann solver \citep{Toro2009}. We set the simulation box in $(x, y, z)$ as (-1.28, 1.28)$\;$Mm $\times$ (1, 30)$\;$Mm $\times$ (-1.28, 1.28)$\;$Mm, where $y=1\;$Mm corresponds to 
$500\;$km above the photosphere. In the numerical simulations, we use a uniform grid within the region below the level given by $y=3.56\;$Mm. 
This region is covered by $256 \times 256 \times 256$ grid points, which leads to a spatial resolution of $10\;$km there. Above this region, i.e.\ in the corona, we implement a divided into $128$ cells stretched vertically grid 
whose cells sizes grow with increasing $y$. As a result of that, any upward propagating signal becomes strongly diffused in the upper layers. This allows us to avoid significant reflections from the upper boundary. At the top, bottom, and sides, we set all the plasma quantities equal to their equilibrium values, given by Eqs.~(14)-(17) and ${\bf V_{\rm i,n}}={\bf 0}$. 

To excite the torsional Alfv\'en waves at the bottom boundary, we set the periodic azimuthal driver in
ion and neutral velocities, which written in Cartesian coordinates takes the form 
\begin{equation}
\begin{split}
& [V_{{\rm i}\,x},V_{{\rm i}\,y},V_{{\rm i}\,z}]=[V_{{\rm n}\,x},V_{{\rm n}\,y},V_{{\rm n}\,z}]=\\
& A_{V} \left[ -z, 0, x \right] \, {\rm exp} \left( - \frac{x^2 + z^2}{w^2}\right) \, {\rm sin}\left( \frac{2 \pi}{P_{\rm d}} t\right) \, , \end{split}
\end{equation}
%
where 
$A_{V}=1\;$km s$^{-1}$ is the effective 
amplitude, $w=0.1\;$Mm the width and $P_{\rm d}$ the period of the driver which we set equal to 30$\;$s. 

\subsection{Numerical results}

Figure~3 shows the spatial profiles of 
azimuthal ion velocity component, 
$V_{{\rm i}\, \theta}(r,y,t)$, with $r=\sqrt{x^2+z^2}$, at three instants of time, namely at $t = 980\;$s, $t = 990\;$s and $t = 1000\;$s (from left to right) for $P_{\rm d}=30\;$s, $A_{V}=1\;$km s$^{-1}$, and $w=100\;$km. 
The torsional driver excites Alfv\'en waves that propagates upwardly along the 
expanding solar magnetic-flux tube. The 
amplitude of these waves grows exponentially 
with the distance at a rate that equals to the pressure-scale height \citep[e.g.,][]{Kuzma2018}, however this effect is 
reduced 
by collisional damping. 
This fact has a two-fold effect on the overall picture. First, less energy is thermalized at the lower layers, and second, more energy can be transferred to the upper layers and contribute to the plasma heating there. The pressure scale-height amplitude growth becomes dominant in the upper chromosphere and above the transition region. At the lower chromospheric heights, the ion-neutral coupling 
is already weak enough for neutrals to depart from the evolution of the ions. It was shown in previous studies on two-fluid acoustic waves \citep{Kuzma2019} that even in non-magnetic case both ion and neutral fluids start to decouple in the chromosphere. Here, the  magnetic field is an additional factor amplifying this effect. The dynamic motions of ions are directly altered by the presence of magnetic field. Obviously, the latter has no such influence on neutrals and, as result, they get extra freedom to propagate in the radial direction. The ion-neutral drift increases, and, as it will be shown later, starts playing a significant role in the process of plasma heating according to Eqs.~(11)-(12).  

The top panel of Fig.~4 illustrates a time-distance plot of the azimuthal component of the ion velocity collected $100\;$km out of the flux-tube center, $V_{{\rm i}\, \theta}(r=0.1\;$Mm). Note that only part of the timeline for $t>10^3\;$s is displayed, namely the period after the transient phase and when the quasi-stationary state is reached. We can see that, although in the chromosphere the torsional Alfv\'en waves are partially damped in process of ion-neutral collisions as 
ionized plasma and neutrals decouple with height, they propagate higher up, ultimately reaching the transition region and corona above (compare with Fig.~3). This stays in an agreement with recent findings of 
\cite{Soler2019}. 
The transverse MHD waves (i.e., Alfv\'en and kink waves) takes part in the process of transport of the energy into and through the upper 
layers of the solar atmosphere. This energy, if thermalized, 
can play a significant role in the heating of the chromosphere and corona \citep[e.g.,][]{Dwivedi2010,Sokolov2013,Murawski2016b,Srivastava2017}. Note that even small ion-neutral drift multiplied by friction coefficient, $\alpha_{c}$, may result in noticeable frictional heating (Eqs. 11 - 12), and thus ion-neutral collisions may play a role in thermalizing energy carried by Alfv\'en waves. The middle panel of Fig.~4 illustrates the corresponding time-distance plot of the Alfv\'en waves ion energy flux,  
described by the following formula \citep{Mathioudakis2013}:
\begin{equation}
\begin{split}F_{E_{{\rm i},\theta}}=\varrho_{\rm i} c_{\rm A}V_{{\rm i}, \theta} \, .
\end{split}
\end{equation}
As we can see, the torsional Alfv\'en waves carry along the flux-tube an energy flux of about $10^4\;$erg cm$^{-2}$ s$^{-1}$. This value falls off with height to about $10^3\;$erg cm$^{-2}$ s$^{-1}$ right above the transition region due to abrupt drop in ion density, $\varrho_{\rm i}$ (not shown). 

As we discussed in Section~2, one of the possible mechanism of dissipation of the carried energy is frictional heating resulting from ion-neutral collisions. The bottom panel of Fig.~4 illustrates the ion-neutral drift, $\log|V_{{\rm i}}-V_{{\rm n}}|(r=0.1\;$Mm) with the $-\theta$ 
component taken into account, thus for Alfv\'en 
waves. From Eqs.~(11)-(12), we infer that the frictional heating of the plasma by torsional Alfv\'en waves is dominant in the chromosphere and significantly grows with height. 
Although Alfv\'en waves are present in the solar corona, frictional heating cannot contribute effectively to coronal heating due to the low abundance of neutrals, as the plasma becomes essentially fully-ionized there. 


\begin{figure}
	\begin{center}
		\mbox{
		 \includegraphics[scale=0.43]{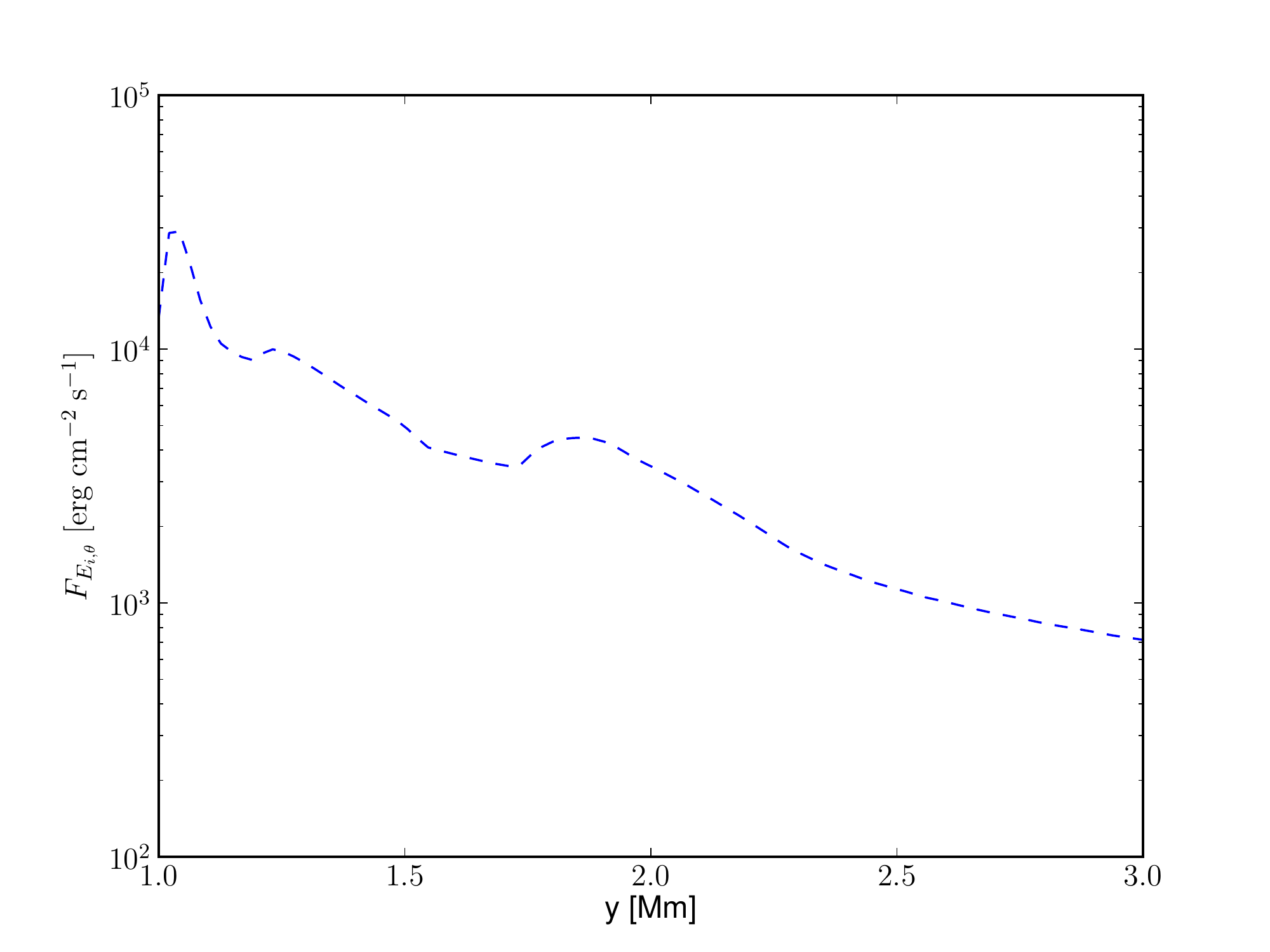}}
		\caption{The maximum value of ion energy flux, $F_{E_{i}}$ carried by Alfv\'en waves 
		versus height, $y$.}
		\label{fig:solar_profiles}
	\end{center}
\end{figure}

Figure~5 reveals the maximum value of ion-neutral drift, $|V_{{\rm i}\, \theta}-V_{{\rm n}\, \theta}|$ versus height, $y$ (top) and time-distance evolution of the relative ion temperature, $\delta T_{\rm i}(r=0.1\;$Mm,$t)=T_{\rm i}(r=0.1\;$Mm,$t)-T_{0}$ (bottom). The ion-neutral drift associated with Alfv\'en waves experiences constant growth above the height of $y=1.1\;$Mm and abruptly jump above $y=1.6\;$Mm. We see that the energy dissipated in the process of frictional heating leads to a significant increase of the plasma temperature, mainly in the upper chromosphere just below the transition region. Note that the relative plasma temperature shares many similarities to the heating pattern of driven in the photosphere acoustic waves \citep{Kuzma2019} 
and 2.5D Alfv\'en waves \citep{Kuzma2020}. 

\begin{figure}
	\begin{center}
%
		\mbox{
					\hspace{-0.5cm}
		 \includegraphics[scale=0.45]{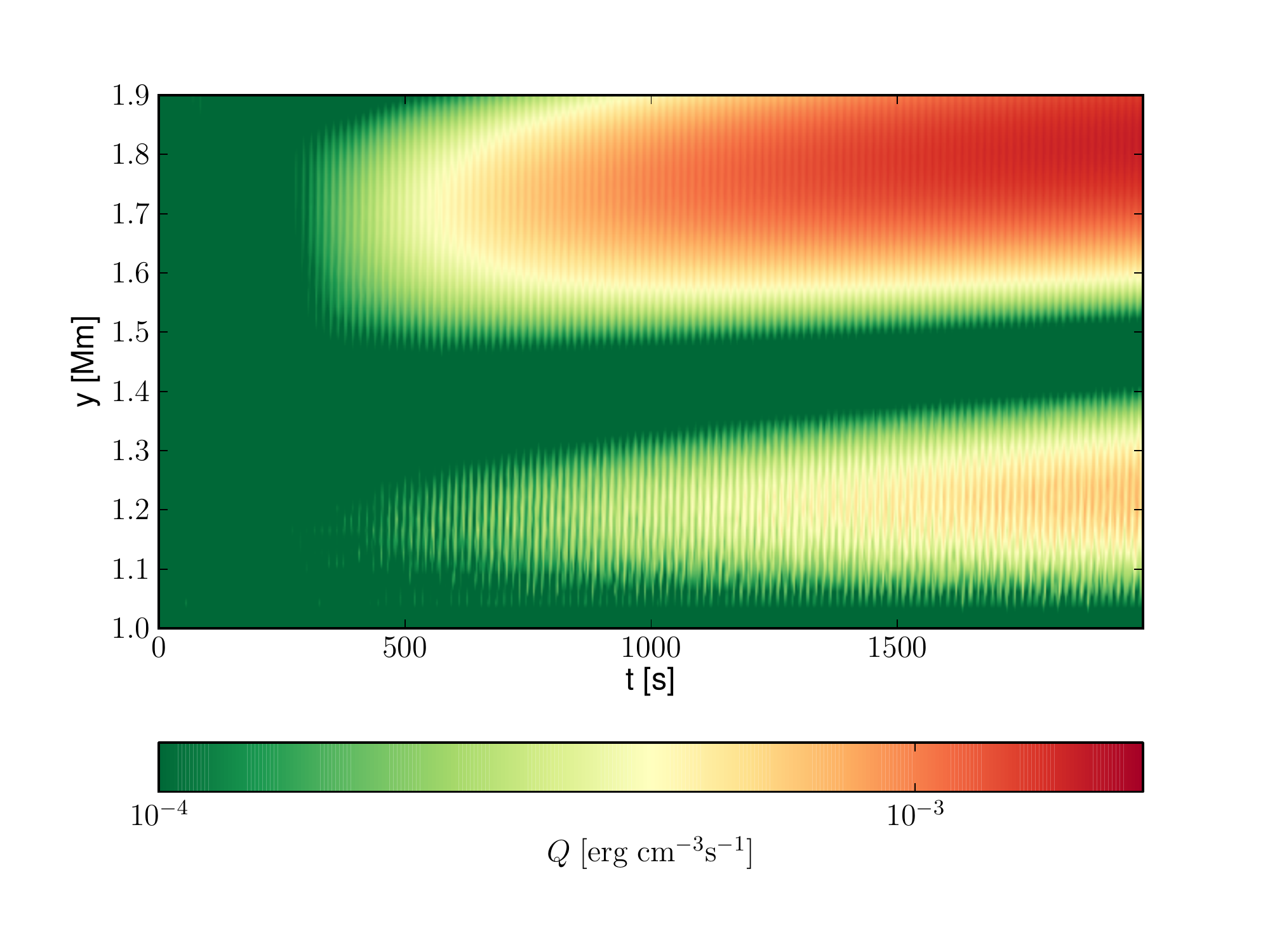}}
\vspace{-0.5cm}
		 \caption{Time-distance plot for the frictional heating rate taken from $Q_{\rm i}$ term.} 
		\label{fig:solar_profiles}
	\end{center}
\end{figure}

Figure~6 shows the height profile of the maximum ion energy flux carried by Alfv\'en waves as described by Eq. (19). Note that after the initial transient phase ($t<50$ s) this profile remains constant in time and depends solely on the driver amplitude. 
We see that Alfv\'en waves in terms of energy transport remain relevant along the entire flux-tube. This energy is partially dissipated in the chromosphere by ion-neutral collisions (compare with Fig.~5, bottom). 
As a result of dissipation, only $10$\% of the initially pumped into system energy reaches the corona as torsional Alfv\'en waves. 
Note that these waves may potentially still dissipate their energy in the corona contributing to coronal heating. 
As we can see, the torsional Alfv\'en waves carry an energy flux of about $10^4\;$erg cm$^{-2}$ s$^{-1}$ in the chromosphere and this value falls off with height to about $10^3\;$erg cm$^{-2}$ s$^{-1}$ right above the transition region. 
In comparison to chromospheric and coronal radiative energy losses for quiet Sun estimated by \citet{Withbroe1977}, the total energy flux carried by Alfv\'en waves is an order of magnitude too small to compensate these losses. Figure 7 illustrates the time-distance plot of collisional heating rate obtained in simulation and taken from Eq. (12). The obtained value of heating rate is also one order of magnitude too small in comparison with the estimated chromospheric cooling rate which in the middle and upper solar chromosphere attains a value of $\sim 10^{-2}$ erg cm$^{-3}$ s$^{-1}$ \citep{Avrett2015}. This scenario may change if we take non-linear Alfv\'en waves into consideration \citep{Kuzma2020}.

%
%
%
\section{Conclusions}
%
By means of numerical simulations we described a partially-ionized solar plasma using two-fluid equations for ions plus electrons and neutral gas, respectively. We coupled these fluids with ion-neutral collisions, and, finally, we generated monochromatic Alfv\'en waves that originate below the chromosphere. We let them propagate along the current-free magnetic field configuration of a magnetic flux-tube and investigated their attenuation and dissipation. We followed and extended the previous studies of \cite{Popescu2019} and \cite{Kuzma2019} by taking into account a 3D model of the solar atmosphere, and the study of \cite{Soler2019} by taking into account non-static, time-varying model of propagating Alfv\'en waves. 

We found that the magnetic flux-tube acts as a guideline for the driven torsional Alfv\'en waves. These waves propagate upwards into the chromosphere and corona. However, only part of the carried wave-energy is transferred above the transition region. Part of this energy is thermalized in the chromosphere in the process of collisional heating. 
We found that although the chromospheric plasma temperature rises in time, both the energy carried by two-fluid Alfv\'en waves and collisional heating can compensate the radiative losses only partially and thus these Alfv\'en waves alone are not sufficient to explain the chromospheric heating. We also infer that as a result of a dissipation $10$\% of the initial energy input reaches the corona in form of Alfv\'en waves. 

Our conclusion has important implications on constructing theoretical models of the solar atmosphere. It was showed that the two-fluid effects may play significant, if not crucial role in 
wave processes in the chromosphere and energy transfer between the photosphere and corona. Therefore they should be included into the model. Perhaps even more important, we showed that two-fluid effects are essential not only for short wave periods. 
We note that the only dissipation mechanism considered in this work is ion-neutral collisions which, according to our results, do not produce enough heating to compensate radiative losses in the chromosphere. This agrees with previous findings. However, if additional dissipation mechanisms were included, the heating would probably be larger as their associated heating rates are even greater than those of ion-neutral collisions, which has been explored in a number of studies (see, for instance,  \citealp{Goodman2011,Tu2013,Arber2016,Soler2019}). In particular, magnetic diffusion has been shown to be the dominant heating mechanism in the most part of the chromosphere (essentially at low and medium altitudes). 
Further research is required to determine the feasibility of constructing the realistic 3D model, i.e.\ with thermal conduction, radiative cooling, non-local thermal equilibrium in the chromosphere and magnetic resistivity terms taken into account, sustaining its quasi-equilibrium with two-fluid Alfv\'en waves heating. 

\section*{Acknowledgements}
The JOANNA code was developed at the  Institute of Mathematics of University of M. Curie-Sk{\l}odowska, Lublin, Poland by Darek W\'ojcik. This work was 
supported as part of project funded by 
National Science Centre (NCN) grant nos. 2017/25/B/ST9/00506 and 2020/37/B/ST9/00184. 
SP and BK acknowledges support through the projects
C14/19/089  (C1 project Internal Funds KU Leuven), G.0D07.19N (FWO-Vlaanderen), SIDC Data Exploitation (ESA Prodex-12). This project (EUHFORIA 2.0) has received funding from the European Union’s Horizon 2020 research and innovation programme under grant agreement No 870405.

\section*{Data Availability Statement}
The data underlying this article will be shared on reasonable request to the corresponding author.

\bibliographystyle{mnras}
\bibliography{draft.bib}
\end{document}